\documentclass[Times,twocolumn,tighten,deluxetable]{aastex7}

\usepackage{graphicx}
\usepackage{rotating}
\usepackage{lineno}
\usepackage{multirow}
\usepackage{enumerate}

\submitjournal{ApJ}

\shorttitle{X-ray Polarization of IGR J17091-3624}
\shortauthors{Debnath et al.}

\begin{document}

\title{Detection of X-ray Polarization in the Hard State of IGR J17091-3624: Spectro-Polarimetric Study with IXPE and NuSTAR Data}

\correspondingauthor{Dipak Debnath}
\email{dipakcsp@gmail.com}

\author[0000-0003-1856-5504]{Dipak Debnath}
\affiliation{Institute of Astronomy, National Tsing Hua University, Hsinchu 300044, Taiwan}
\affiliation{Institute of Astronomy Space and Earth Science, P 177, CIT Road, Scheme 7m, Kolkata 700054, India}
\email{dipakcsp@gmail.com}

\author[0009-0004-5915-3789]{Subham Srimani}
\affiliation{Institute of Astronomy, National Tsing Hua University, Hsinchu 300044, Taiwan}
\email{subhamsrimani2023@gmail.com}

\author[0000-0002-5617-3117]{Hsiang-Kuang Chang}
\affiliation{Institute of Astronomy, National Tsing Hua University, Hsinchu 300044, Taiwan}
\affiliation{Department of Physics, National Tsing Hua University, Hsinchu 300044, Taiwan}
\email{hkchang@mx.nthu.edu.tw}


\begin{abstract}

The class-transition Galactic X-ray binary IGR~J17091--3624 was simultaneously monitored by the \textit{IXPE} and \textit{NuSTAR} satellites. 
We present a detailed spectro-polarimetric study of the source using data from both satellites covering the period from March~7--10, 2025.
A polarimetric analysis in the $2$--$8$~keV band using a model-independent method reveals a significant detection of polarization degree (PD) 
of $(11.3\pm2.35)\%$ at a polarization angle (PA) of $82^\circ.7\pm5^\circ.96$ (significant at $>4\sigma$). The model-dependent polarization 
analysis using the \texttt{polconst} and \texttt{polpow} models yields consistent values of PD and PA. In both methods, an energy-dependent 
increasing trend of PD is observed. In the $6$--$8$~keV band, a maximum PD of $(29.9\pm8.46)\%$ (significant at $>3\sigma$) is detected at 
a PA of $88^\circ.0\pm8^\circ.15$ ($>3\sigma$).
The joint spectral analysis using \textit{IXPE} and \textit{NuSTAR} data in the $2$--$70$~keV band was performed with four different sets 
of phenomenological and physical models. Our results indicate a strong dominance of non-thermal photons originating from a `hot' Compton cloud, 
suggesting that the source was in a hard spectral state. Spectral fitting with the physical {\fontfamily{qcr}\selectfont kerrbb} and 
{\fontfamily{qcr}\selectfont TCAF} models provides an estimate of the black hole mass $M_{\rm BH} = 14.8^{+4.7}_{-3.4}~M_\odot$ and 
dimensionless spin parameter $a^* \sim 0.54$.
The requirement of a higher hydrogen column density in the spectral fit of the second \textit{NuSTAR} observation is attributed to 
the obscuration of non-thermal photons during the dip phase, likely caused by the presence of wind accreted from the companion star.

\end{abstract}


\section{Introduction}

The energy spectrum of black holes mainly consists of two components: a thermal multi-color disk blackbody (DBB) and a non-thermal power-law (PL). 
The DBB component is believed to originate from the standard Keplerian disk \citep{NT73, SS73}, while the PL component is thought to arise from a hot, 
electron-filled Compton cloud or corona \citep{ST80, ST85}. Soft photons from the Keplerian disk are upscattered within the corona and emerge as hard 
PL photons. Jets and outflows are believed to be launched from the same corona \citep{C99}, and the emitted radiation becomes polarized depending on 
the geometry, optical depth, temperature, and inclination angle of the system \citep{Connors80, ST85, Chauvin18, Krawczynski22}. 
The black hole (BH) spin also significantly impacts the degree of polarization, in addition to the above accretion and geometrical properties. 
Thus, the polarization properties—such as polarization degree (PD) and polarization angle (PA)—depend on several factors, including 
the geometrical configuration of the corona, accretion flow dynamics, and BH spin. 

Observations of polarization in BHs and other accreting systems are not new and have been extensively studied in the past, from X-rays to 
$\gamma$-rays \citep[see for e.g.,][]{Long80, Laurent11, Rodriguez15, Vadawale18}. The launch of the 
\textit{Imaging X-ray Polarimetry Explorer} (\textit{IXPE}) in December 2021 significantly advanced our understanding of high-energy astrophysical 
phenomena by enabling unprecedented X-ray polarization measurements. This allows for a more refined understanding of accretion–ejection processes and 
the geometrical configuration of accreting sources through X-ray polarimetric studies. The high-quality, long-duration observations from \textit{IXPE} 
can robustly measure PD and PA, i.e., the polarization fraction and the direction of the electric field vector in the polarized emission.

Recently, polarization has been detected in many BH candidates (BHCs) using \textit{IXPE} data. Evolution of PD in different spectral 
states of the same BHC has also been observed. \citet{Jana24} reported PD variation from 4\% to 2.5\% during the transition from the hard state (HS) 
to the soft state (SS) in Cyg~X-1. In the HS of Swift~J1727.8-1613, a PD of $\sim 3$-$4$\% has been observed \citep{Ingram23, Veledina23, Podgorny24}, 
while in the SS, a rapid decrease in PD to $\sim 0.5$\% was noted \citep{Svoboda24a}. A PD of $\sim 6.5$-$10$\% was detected in 4U~1630--47 
\citep{Kushwaha23a, Rodriguez23, Rawat23, Ratheesh24}. LMC~X-3 showed a PD of $\sim 3$-$4$\% in the SS \citep{Svoboda24b, Majumder24}, 
while a PD of $\sim 2$\% was found in 4U~1957+115 \citep{Kushwaha23b, Marra24}. A large PD of $\sim 25$\% has been observed in the high-mass 
X-ray binary Cyg~X-3 \citep{Veledina24}. In persistent low mass X-ray binary Cyg~X-1, a large variation in PD from $2.4$ to $75$\% was observed 
across multi-satellite studies from X-rays to $\gamma$-rays \citep{Long80, Jourdain12, Rodriguez15, Chauvin18, Chattopadhyay23, Jana24}. 

IGR~J17091--3624, one of the two renowned class-transition Galactic low-mass X-ray binaries, was discovered during an \textit{INTEGRAL} 
Galactic Center Deep Exposure \citep{Kuulkers03} in 2003. The subsequent 2011 outburst of the source was particularly remarkable, 
during which it exhibited eight ($\mu$, $\nu$, $\lambda$, $\kappa$, $\rho$, $\beta$, $\chi_2$, $\chi_3$) out of the fourteen different types 
of light curves, including the heart-beat ($\rho$) class originally observed in GRS~1915+105 \citep[][and references therein]{Belloni00, Pal13, Pal15}. 

During the rising phase of the 2011 outburst, prior to entering a prolonged intermediate state where it displayed various classes 
of variability, IGR~J17091-3624 showed similar outburst characteristics to other canonical transient BHCs. 
In the onset phase, the evolution of low-frequency quasi-periodic oscillations (LFQPOs) was reported by \citet{Iyer15}. The source 
has been extensively studied during its past outbursts. A wide range of values have been reported for its BH mass ($M_{\rm BH}$) 
and distance ($D$): $M_{\rm BH} < 3~M_\odot$ \citep{Altamirano11}, $<5~M_\odot$ \citep{Rao12}, $\sim 15~M_\odot$ \citep{Altamirano12}, 
and $8.7$-$15.6~M_\odot$ \citep{Iyer15}; and $D \sim 11$-$17$~kpc \citep{Rodriguez11} to $>20$~kpc \citep{Rao12}. The spin of the black 
hole has been estimated to be $a^* > 0.53$ and the inclination angle has been reported as $i > 53^\circ$ by \citet{Rao12}. 
Notably, the source also exhibited a high-frequency QPO (HFQPO) at 66~Hz, similar to GRS~1915+105 \citep{Altamirano12}. Many comparative 
studies between IGR~J17091-3624 and GRS~1915+105 have been conducted to understand the similarities and differences between the two sources 
\citep[e.g.,][]{Capitanio12, Pal15, Katoch21, Banerjee22}.

Following the 2011-13 outburst, the source underwent two additional outbursts in 2016 and 2022. The most recent outburst was detected 
by \textit{INTEGRAL} on 2025 February 7 \citep{Rodriguez25}. A hard spectral state with a photon index of $\sim 1.47$ and LFQPOs at 
$\sim 0.2$~Hz was observed with \textit{NICER} \citep{Vincentelli25}. A preliminary X-ray polarization detection in the 
2--8~keV band with a PD of $(9.3 \pm 1.8)$\% and PA of $82^\circ \pm 5^\circ$ was reported using \textit{IXPE} data \citep{Parra25}. 
\citet{Ewing25} also reported detection of similar PD ($(9.1 \pm 1.6)$\%) and PA ($83^\circ \pm 5^\circ$).

In this \textit{paper}, we present a spectro-polarimetric study of IGR~J17091--3624 using observations from the \textit{IXPE} 
and \textit{NuSTAR} satellites. Polarization measurements are performed using both model-independent and model-dependent approaches. 
A broadband spectral study is also carried out to investigate the accretion flow properties and to estimate the BH mass, 
spin, and inclination angle. The paper is organized as follows: \S 2 describes the observations, data reduction, and analysis 
procedures. In \S 3, we present the results, while \S 4 discusses our findings and draws conclusions.

\section{Observation and Data Analysis}

We analyze a $163$~ks observation (March 7–10, 2025; MJD = 60741.31–60744.77) from the {\it IXPE} satellite and two overlapping 
{\it NuSTAR} observations: $21$~ks (March 7–8, 2025; MJD = 60741.58–60742.04, hereafter Nu1) and $19$~ks (March 8–9, 2025; 
MJD = 60742.98–60743.43, hereafter Nu2).

\begin{figure*}
  \centering
    \includegraphics[angle=0,width=16.0cm,keepaspectratio=true]{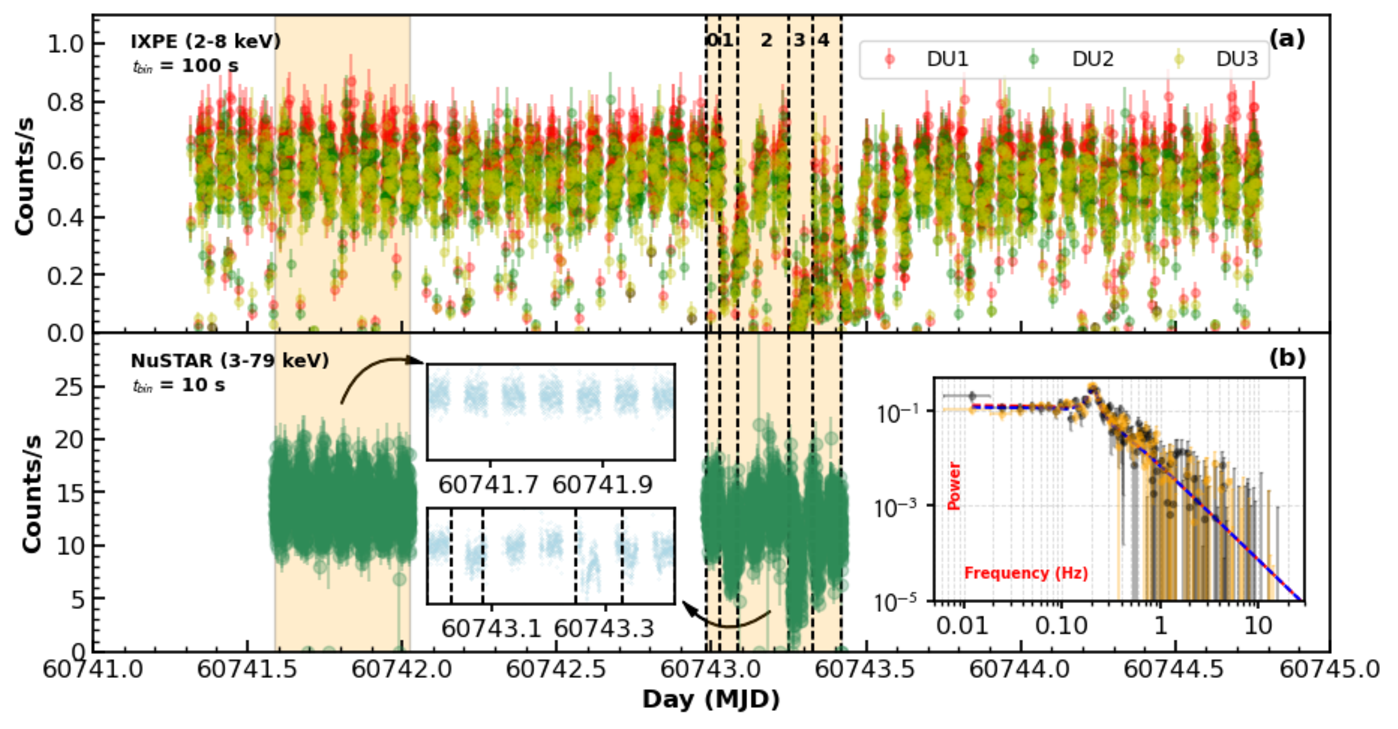}
      \caption{(a) Light curve of {\it IXPE} in the $2$–$8$~keV band ($100$~s time bin) for all three DUs. (b) Light curve of {\it NuSTAR}/FPMA 
	in the $3$–$70$~keV band ($10$~s time bin). The power density spectra from $0.01$~s time-binned {\it NuSTAR} light curves for Nu1 
	and Nu2 (orange and black points) are fitted with a Lorentzian model (blue and red dashed lines), revealing a distinct QPO at 
	$0.21\pm0.004$~Hz. The vertical black dashed lines mark five regions of the Nu2 observation.}
\label{fig_lc}
\end{figure*}

\subsection{Data Analysis}  

We follow standard data reduction procedures for both {\it IXPE}\footnote{\url{https://ixpeobssim.readthedocs.io}} and 
{\it NuSTAR}\footnote{\url{https://heasarc.gsfc.nasa.gov/docs/nustar/analysis}}. All data analysis is performed using the online
platform SciServer\footnote{\url{https://www.sciserver.org}}, employing the latest version (v6.34) of the HEASoft package from HEASARC.

\subsubsection{IXPE}
The {\it IXPE} archival data (ObsID 04250201) is processed using the latest version (v31.0.3) of the {\fontfamily{qcr}\selectfont ixpeobssim} 
software \citep{Baldini22}. Level 2 event files from all three detector units (DUs) are used in our analysis. Light curves are extracted using 
the {\fontfamily{qcr}\selectfont xpbin} task \citep{Kislat15}. Source and background event files are generated using the 
{\fontfamily{qcr}\selectfont xpselect} task, employing region files created with DS9: a circular source region of $50''$ radius centered at the 
source coordinates, and an annular background region with inner and outer radii of $180''$ and $240''$, respectively. The I, Q and U Stokes 
parameter files ({\fontfamily{qcr}\selectfont PHA1, PHA1Q, PHA1U}) and {\fontfamily{qcr}\selectfont PCUBE} algorithm based files are also 
generated using {\fontfamily{qcr}\selectfont xpbin}. For spectral analysis, we rebin the {\fontfamily{qcr}\selectfont PHA1} spectra to ensure 
at least $100$ counts per bin using the {\fontfamily{qcr}\selectfont GRPPHA} task.

\subsubsection{NuSTAR}
The {\it NuSTAR} archival data for Nu1 (ObsID 81002342008) and Nu2 (ObsID 81002342010) are reduced using the 
{\fontfamily{qcr}\selectfont NuSTARDAS} software (v2.1.4a). Cleaned event files are produced with the {\fontfamily{qcr}\selectfont nupipeline} 
task using the latest calibration files. Source spectra are extracted from a circular region of $50''$ radius centered at the source coordinates, 
while background spectra are extracted from an annular region ($180''$–$240''$) defined using DS9. The {\fontfamily{qcr}\selectfont nuproducts} 
task is used to generate source spectra, auxiliary response files (ARF), and response matrix files (RMF). The extracted spectra are then 
rebinned to have at least $100$ counts per bin using {\fontfamily{qcr}\selectfont GRPPHA}.

\section{Results}

\subsection{Light Curves and QPOs}
The variations in the {\it IXPE} and {\it NuSTAR} light curves are shown in Fig.~\ref{fig_lc}. During the second {\it NuSTAR} observation, 
large fluctuations in X-ray intensities are observed across all three {\it IXPE} detector units (DUs) as well as in {\it NuSTAR}/FPMA. 
The light curves generated with a time bin of $0.01$~s for Nu1 and Nu2 are used to construct power density spectra, which show 
prominent QPOs at frequencies of $0.207 \pm 0.004$~Hz and $0.210 \pm 0.004$~Hz, respectively 
(see insets in the bottom panel of Fig.~\ref{fig_lc}).

\begin{figure}
  \centering
    \includegraphics[angle=0,width=9.0cm,keepaspectratio=true]{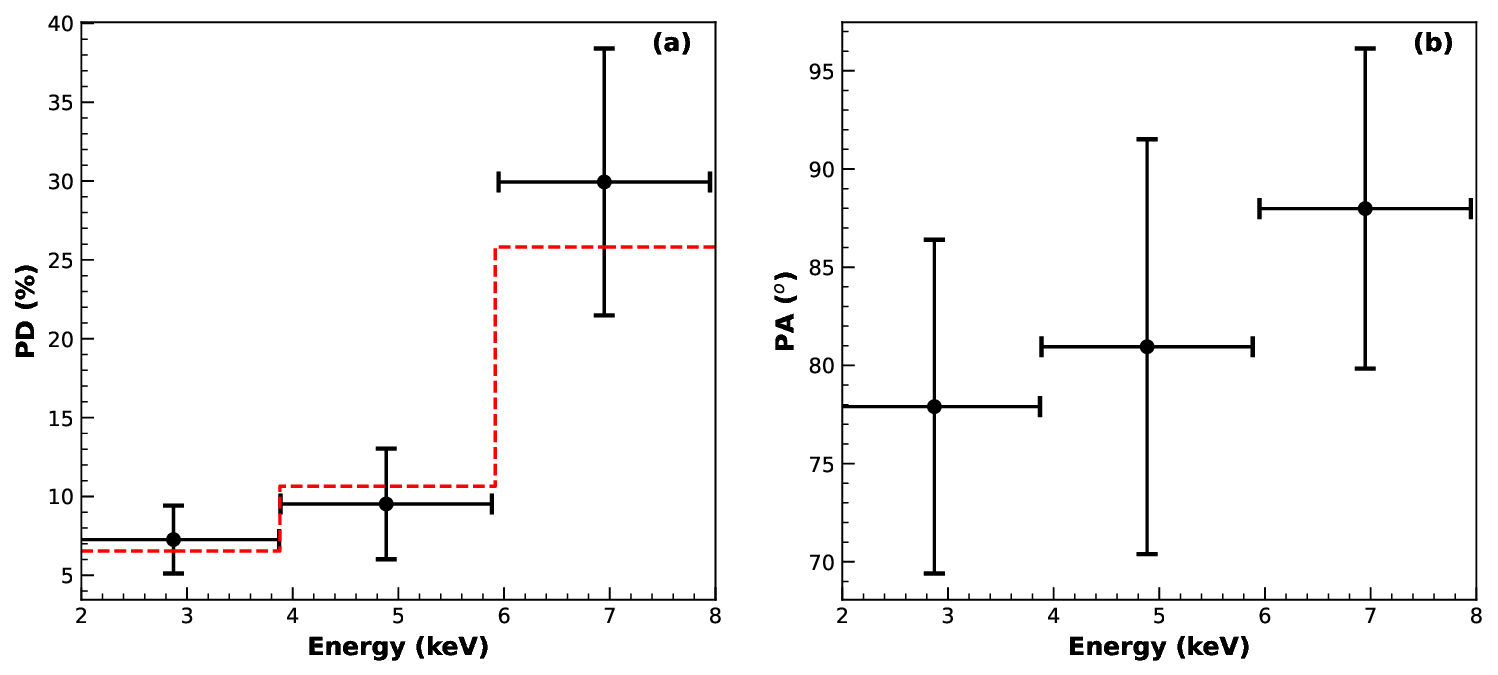}
    \includegraphics[angle=0,width=9.0cm,keepaspectratio=true]{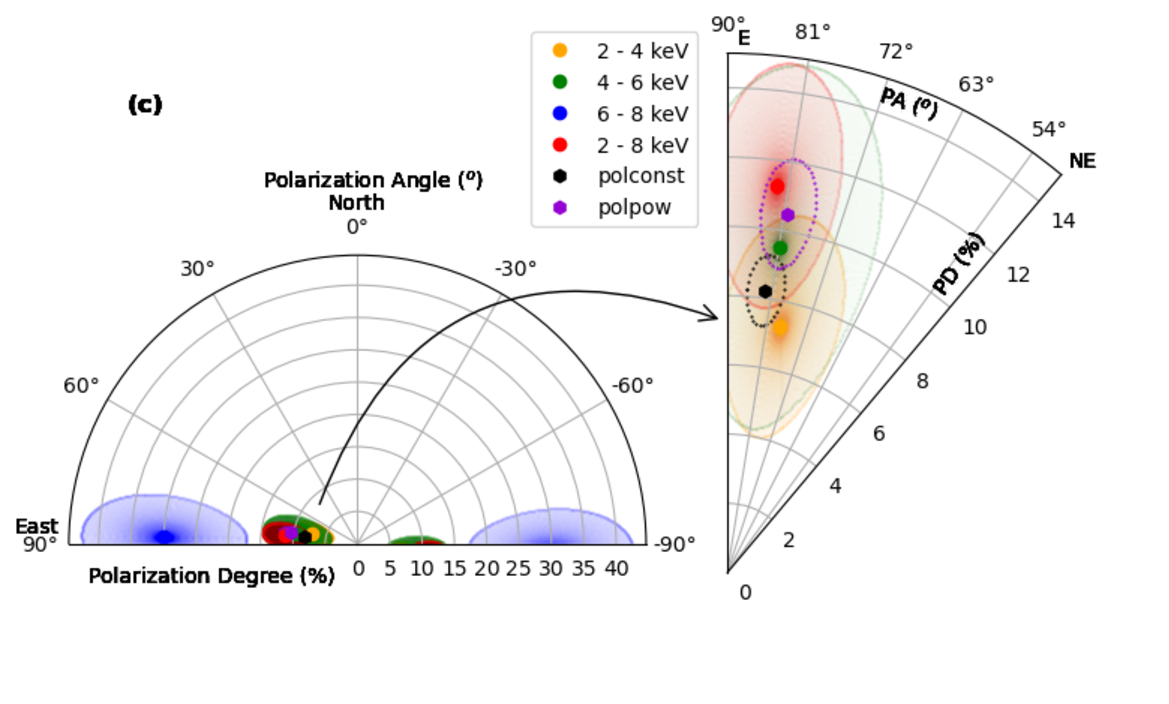}
     \caption{Energy-dependent (a) polarization degree (PD), and (b) polarization angle (PA) in the $2$–$8$~keV range, shown with $1\sigma$ 
	confidence intervals. The red dashed line in panel (a) represents the minimum detectable polarization (MDP) at 99\% confidence level. 
	(c) Confidence contours of energy-resolved polarization parameters of IGR~J17091$-$3624 obtained using a model-independent PCUBE 
	algorithm in the $2$–$4$, $4$–$6$, $6$–$8$, and $2$–$8$~keV bands of {\it IXPE}. Shaded regions bounded by dotted contours represent 
	$1\sigma$ confidence levels. The best-fit polarization values from model-dependent spectral fits using \texttt{polconst} 
	and \texttt{polpow} models in the $2$–$8$~keV band are also indicated along with their respective confidence contours. }
\label{fig_pdpa}
\end{figure}

\subsection{Model-Independent Polarimetric Study}
We use the model-independent \texttt{PCUBE} algorithm of the \texttt{xpbin} task to generate polarization 
cube FITS files in the $2$–$4$, $4$–$6$, $6$–$8$, and $2$–$8$~keV energy bands of {\it IXPE}. From these files, we obtain the following 
polarization parameters: polarization degree (PD), polarization angle (PA), normalized Q-Stokes parameter ($Q/I$), normalized U-Stokes 
parameter ($U/I$), minimum detectable polarization (MDP), and detection significance (SIGNIF). The energy-dependent variations of PD 
and PA for the three DUs are shown in Fig.~\ref{fig_pdpa}(a-b). In the same figure, variations of PD with PA and their corresponding 
$1\sigma$ confidence contours, are plotted for the aforementioned energy ranges. The numerical values of these parameters are listed in 
Table~\ref{table_pol}.

A significant PD of $(11.3 \pm 2.35)\%$ at a PA of $(82.7 \pm 5.95)^\circ$ is observed in the $2$–$8$~keV band, with a confidence level 
exceeding $4\sigma$. This PD is measured from north to east in the sky coordinate system (see Fig.~\ref{fig_pdpa}(c)). Furthermore, the 
PD exhibits a general increasing trend from lower to higher energies, evolving from $(7.27 \pm 2.16)\%$ to $(9.53 \pm 3.51)\%$ to 
$(29.9 \pm 8.46)\%$ at PAs of $(77.9 \pm 8.50)^\circ$, $(81.0 \pm 10.6)^\circ$, and $(88.0 \pm 8.15)^\circ$ in the $2$–$4$, $4$–$6$, and 
$6$–$8$~keV bands, respectively. The 99\% confidence MDP values further confirm the presence of significant polarization in IGR~J17091-3624. 

The normalized Q- and U-Stokes parameters are found to be $(-10.9 \pm 2.35)\%$ and $(2.84 \pm 2.35)\%$, respectively, in the $2$–$8$~keV band. 
As of PD, a similar energy-dependent trend is observed in the Q-Stokes parameter, increasing from $(-6.63 \pm 2.15)\%$ to $(-29.9 \pm 8.45)\%$ 
across the energy bands. However, the U-Stokes parameter remains confined within a narrow range of $(2.11–2.96)\%$.

\begin{table}
\addtolength{\tabcolsep}{-3.5pt}
\centering
\caption{Polarimetric results of IGR~J170091-3624 in different energy bands. Here, $PD$, $Q/I$, $U/I$, and $MDP$ stand for percentage of 
polarization degree, normalized Q-Stokes parameter, normalized U-Stokes parameter, and minimum detectable polarization in 99\% confidence, 
respectively. $PA$ is the polarization angle (in degrees), and $SIGNIF$ is the detection significance (in $\sigma$).
$F_{DBB}$, and $F_{PL}$ refer disk blackbody and powerlaw model fluxes (in units of $10^{-11}~erg~cm^{-2}~s^{-1}$) respectively 
obtained from broadband ($2$–$70$~keV) spectral fitting using combined {\it IXPE} and {\it NuSTAR} data for Nu1 and Nu2.}
\small
\vskip -0.3cm
\begin{tabular}{lccccc}
 \hline
Parameters        & $2$–$4$~keV       & $4$–$6$~keV        &  $6$–$8$~keV       & $2$–$8$~keV       \\
 \hline
	\multicolumn{5}{c}{Model Independent Result} \\
PD       	  &  7.27$^{\pm2.16}$ &  9.53$^{\pm3.51}$  &  29.9$^{\pm8.46}$  &  11.3$^{\pm2.35}$ \\
PA($^\circ$)   	  &  77.9$^{\pm8.50}$ &  81.0$^{\pm10.6}$  &  88.0$^{\pm8.15}$  &  82.7$^{\pm5.96}$ \\
Q/I(\%)       	  & -6.63$^{\pm2.15}$ & -9.05$^{\pm3.51}$  & -29.9$^{\pm8.45}$  & -10.9$^{\pm2.35}$ \\
U/I(\%)       	  &  2.98$^{\pm2.16}$ &  2.96$^{\pm3.51}$  &  2.11$^{\pm8.51}$  &  2.84$^{\pm2.35}$ \\
MDP(\%)       	  &  6.54             &  10.7              &   25.8             &  7.12             \\
SIGNIF ($\sigma$) &  2.93             &  2.24              &   3.11             &  4.43             \\
 \hline
	\multicolumn{5}{c}{Model Dependent Result} \\
$PD_{polconst}$   & ---                & ---               & ---                & 7.92$^{\pm1.03}$   \\
$PA_{polconst}$   & ---                & ---               & ---                & 82.2$^{\pm3.73}$   \\
$PD_{polpow}$     & 2.59$^{\pm0.39}$   & 8.83$^{\pm1.33}$  & 20.1$^{\pm3.04}$   & 10.5$^{\pm1.59}$   \\
$PA_{polpow}$     & 75.8$^{\pm3.95}$   & 81.1$^{\pm4.23}$  & 84.8$^{\pm4.42}$   & 80.6$^{\pm4.20}$   \\
 \hline
\multirow{2}{*}{$F_{PL}/F_{DBB}$} &  6.04/1.00  & 5.84/0.045  & 4.97/0.0010      & 16.85/1.04        \\
                  &  5.41/1.58        & 5.92/0.058         & 5.21/0.0009         & 16.56/1.64        \\
\hline
\end{tabular}
\label{table_pol}
\end{table}

\subsection{Spectro-Polarimetric Study}
We further perform spectro-polarimetric analysis to examine the model-dependent polarization properties of IGR~J17091-3624. For this, we generate 
Stokes I, Q, and U spectral files in the $2$–$8$~keV band for both the source and background using the \texttt{PHA1, PHA1Q,} 
and \texttt{PHA1U} algorithms of the \texttt{xpbin} task. These Stokes spectra from all three DUs 
(total of nine files) are fitted simultaneously using two model sets such as $(1)$ {\fontfamily{qcr}\selectfont tbabs$\otimes$polconst$\otimes$powerlaw}, 
$(2)$ {\fontfamily{qcr}\selectfont tbabs$\otimes$polpow$\otimes$powerlaw} in XSPEC \citep{Arnaud96}. 

From the best fit using model set (1) ($\chi^2_{\mathrm{red}} = 1.11$), the \texttt{polconst} model directly provides the 
polarization parameters: a PD of $(7.92 \pm 1.03)\%$ at a PA of $(82.16 \pm 3.73)^\circ$. However in the model set (2) fit, the PD and PA are 
computed from the \texttt{polpow} model parameters: $A_{\mathrm{norm}}$, $A_{\mathrm{index}}$, $\psi_{\mathrm{norm}}$, and 
$\psi_{\mathrm{index}}$. The energy dependence of PD and PA is described by:
\[
PD(E) = A_{\mathrm{norm}} \times E^{-A_{\mathrm{index}}}, \quad
PA(E) = \psi_{\mathrm{norm}} \times E^{-\psi_{\mathrm{index}}}.
\]

While fitting, we fix the indices of the \texttt{polpow} model at their best-fit values, $A_{\mathrm{index}} = -2.48$ and 
$\psi_{\mathrm{index}} = -0.13$, as their uncertainties could not be reliably constrained. From the best fit ($\chi^2_{\mathrm{red}} = 1.12$), 
we obtain $A_{\mathrm{norm}} = (1.59 \pm 0.24) \times 10^{-3}$ and $\psi_{\mathrm{norm}} = 65.85 \pm 3.43$. These yield a PD of $(10.51 \pm 1.59)\%$ 
and a PA of $(80.56 \pm 4.20)^\circ$ in the $2$–$8$~keV band. 
To explore the energy dependence of polarization, we also estimate PD and PA in individual energy bands using the same \texttt{polpow} 
model fitted parameters. Similar to the model-independent measurements, a monotonic increase in PD is observed, from 
$(2.59 \pm 0.39)\%$ to $(20.1 \pm 3.04)\%$ across three energy bands within $2$–$8$~keV.

In the $2$–$8$~keV band, the model-independent analysis yields PD and PA values of $(11.3 \pm 2.35)\%$ and $(82.7 \pm 5.96)^\circ$, which closely match 
the results obtained from the \texttt{polpow} model in set (2). The PD and PA derived from the \texttt{polconst} 
model in set (1) are also consistent within their respective uncertainties. In Fig.~\ref{fig_pdpa}, the polarization parameters obtained from the spectral 
fits are plotted alongside the model-independent PD and PA values.

\begin{figure}
  \centering
    \includegraphics[angle=0,width=9.0cm,keepaspectratio=true]{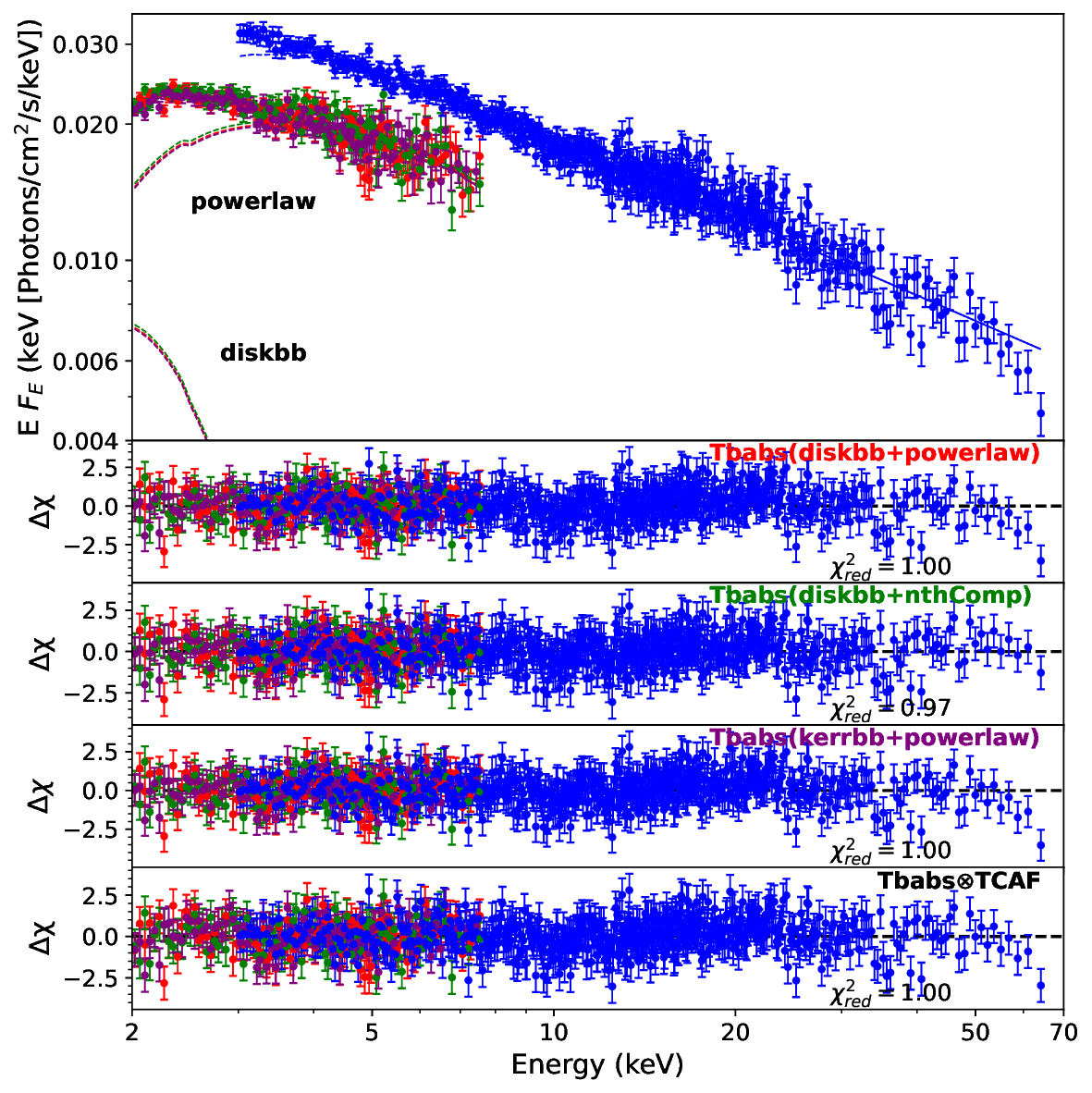}
	\caption{Broadband spectrum using combined IXPE ($2-8$~keV) and first NuSTAR (Nu1 in $3$--$79$ keV) data of IGR~J17091-3624 are 
        fitted with different set of spectral models:
        (a) $constant \otimes tbabs (diskbb + powerlaw)$,
        (b) $constant \otimes tbabs (diskbb + nthComp)$,
        (c) $constant \otimes tbabs (kerrbb + powerlaw)$, and
        (d) $constant \otimes tbabs \otimes TCAF$.}
\label{fig_spec}
\end{figure}

\begin{table*}
	\centering
	\vskip 0.0cm
	\addtolength{\tabcolsep}{-4.0pt}
	\small
	\caption{Broadband ($2-70$~keV) Spectral Fitted Results using Simultaneous IXPE and NuSTAR Observations} 
	\label{tab:table2}
	\vskip -0.3cm
	\begin{tabular}{lcc|cr||lcc|cr} 
		\hline
  \hline
		& Nu1              & Nu2    &  Nu2$_{r024}$$^\dagger$  & Nu2$_{r13}$$^\dagger$  &    & Nu1           & Nu2   &  Nu2$_{r024}$$^\dagger$  & Nu2$_{r13}$$^\dagger$                \\
  \hline
  \multicolumn{5}{c||}{\textsc{Tbabs(diskbb + powerlaw)}}        & \multicolumn{5}{c}{\textsc{Tbabs(diskbb + nthComp)}}     \\
 $N_{\rm H}$   & 1.87$^{\pm0.25}$  & 2.61$^{\pm0.23}$  & 2.06$^{\pm0.56}$    & 6.03$^{\pm0.48}$       & ~$N_{\rm H}$   & 1.87$^a$          & 2.61$^a$         & 2.03$^a$          & 6.03$^a$            \\
 $kT_{\rm in}$ & 0.44$^{\pm0.02}$  & 0.41$^{\pm0.01}$  & 0.49$^{\pm0.04}$    & 0.39$^{\pm0.02}$       & ~$kT_{\rm in}$ & 0.42$^{\pm0.02}$  & 0.40$^{\pm0.01}$ & 0.45$^{\pm0.02}$  & 0.38$^{\pm0.02}$   \\
 N$_{DBB}$     & 223$^{\pm103}$    & 784$^{\pm102}$    & 3.17$^{\pm1.61}$    & 21.6$^{\pm9.8}$        & ~N$_{DBB}$     & 232$^{\pm61}$     & 936$^{\pm161}$   & 6.41$^{\pm2.16}$  & 22.7$^{\pm9.8}$      \\
 $\Gamma$      & 1.58$^{\pm0.01}$  & 1.53$^{\pm0.01}$  & 1.54$^{\pm0.02}$    & 1.56$^{\pm0.02}$       & ~$\Gamma$      & 1.62$^{\pm0.01}$  & 1.59$^{\pm0.01}$ & 1.59$^{\pm0.05}$  & 1.61$^{\pm0.07}$   \\
 N$_{PL}$      & 0.05$^{\pm0.001}$ & 0.05$^{\pm0.001}$ & 1.2$^{\pm0.05}$$\times10^{-3}$ & 4.2$^{\pm0.3}$$\times10^{-4}$  & ~$kT_e$        & 25.6$^{\pm2.41}$  & 32.8$^{\pm6.32}$ & 24.8$^{\pm4.54}$  & 32.0$^{\pm11.2}$   \\
 $\chi^2/DOF$  & 817/813           &    794/777        & 742/701            & 463/419                & ~$kT_{bb}$     & 0.40$^{\pm0.12}$  & 0.38$^{\pm0.02}$ & 0.46$^{\pm0.07}$ & 0.39$^{\pm0.02}$   \\
 $F_{DBB}$     & 0.15              & 0.19              & 0.045              & 0.059                  & ~N$_{NCOMP}$   & 0.03$^{\pm0.01}$  & 0.03$^{\pm0.001}$& 5.6$^{\pm0.12}$$\times10^{-4}$ & 2.4$^{\pm0.15}$$\times10^{-4}$ \\  
 $F_{PL}$      & 12.0              & 10.9              & 11.06              & 9.58                   & ~$\chi^2/DOF$  & 790/811           & 772/775          & 730/700           & 454/418           \\
 FR$^b$	       & 0.0125            & 0.0174            & 0.0041             & 0.0062                 & ~$F_{DBB}$     & 0.16              & 0.19             & 0.044              & 0.059             \\
	       & ~                 & ~                 & ~                   & ~                      & ~$F_{NCOMP}$   & 11.7              & 10.7             & 10.9               & 9.30              \\
	       & ~                 & ~                 & ~                   & ~                      & ~FR$^b$        & 0.0137            & 0.0178           & 0.0040            & 0.0063             \\
 \hline
 \multicolumn{5}{c||}{\textsc{Tbabs(kerrbb + powerlaw)}}         & \multicolumn{5}{c}{\textsc{Tbabs$\otimes$TCAF}}\\
 $N_{\rm H}$   & 1.87$^a$          & 2.61$^a$                   & 2.03$^a$               & 6.03$^a$                 & ~$N_{\rm H}$   & 1.87$^a$   & 2.61$^a$                  & 2.03$^a$            & 6.03$^a$          \\
 $a$          & 0.54$^{+0.09}_{-0.07}$ & 0.54$^{+0.14}_{-0.25}$ & 0.54$^{\pm0.03}$       & 0.53$^{\pm0.16}$         & ~$\dot{m}_d$  & 0.002$^{\pm0.001}$ & 0.004$^{\pm0.001}$ & 0.005$^{\pm0.001}$  & 0.033$^{\pm0.006}$\\
 $i$          & 56$^a$                 & 56$^a$                 & 56$^a$                 & 56$^a$                   & ~$\dot{m}_h$  & 0.811$^{\pm0.066}$ & 0.804$^{\pm0.018}$ & 0.782$^{\pm0.006}$  & 1.032$^{\pm0.034}$\\ 
 $M_{BH}$     & 14.9$^{+0.4}_{-1.3}$   & 14.6$^{+4.8}_{-3.4}$   & 14.9$^{\pm1.0}$        & 14.7$^{\pm0.6}$          & ~$M_{BH}$     & 14.4$^{\pm1.7}$    & 15.0$^{\pm3.6}$    & 14.9$^{\pm0.4}$     & 14.9$^{\pm0.1}$   \\
 $\dot{M}_d$  & 0.29$^{+0.09}_{-0.06}$ & 0.22$^{+0.12}_{-0.04}$ & 0.48$^{\pm0.03}$       & 0.18$^{\pm0.007}$        & ~$X_s$        & 446$^{\pm17}$      & 460$^{\pm31}$      & 410$^{\pm3.2}$      & 460$^{\pm18}$     \\
 D            & 14$^a$                 & 14$^a$                 & 14$^a$                 & 14$^a$                   & ~$R$          & 3.18$^{\pm0.03}$   & 3.15$^{\pm0.05}$   & 3.07$^{\pm0.02}$    & 3.26$^{\pm0.18}$  \\
 N$_{KERRBB}$ & 0.29$^{+0.11}_{-0.09}$ & 0.70$^{+0.21}_{-0.24}$ & 2.7$^{\pm0.21}$$\times10^{-3}$ & 0.021$^{\pm0.002}$ & ~N$_{TCAF}$   & 0.32$^{\pm0.08}$   & 0.28$^{\pm0.02}$   & 6.5$^{\pm0.15}$$\times10^{-3}$ & 7.3$^{\pm0.43}$$\times10^{-4}$ \\
 $\Gamma$     & 1.58$^{\pm0.01}$   & 1.53$^{\pm0.01}$           & 1.54$^{\pm0.01}$       & 1.56$^{\pm0.01}$         & ~$\chi^2/DOF$ & 816/811            & 828/775            & 743/699            & 470/417           \\
 N$_{PL}$     & 0.05$^{\pm0.002}$   & 0.05$^{\pm0.002}$         & 1.2$^{\pm0.04}$$\times10^{-3}$ & 4.2$^{\pm0.29}$$\times10^{-4}$ & ~$F_{TCAF}$   & 12.1 & 11.1               &  11.08  & 9.69              \\
 $\chi^2/DOF$ & 817/810            & 793/774                   & 741/700                & 463/418                  & ~ & ~ &~ & ~ &\\
 $F_{KERRBB}$ & 0.15               & 0.19                      & 0.048                  & 0.059                    & ~ & ~ &~ & ~ &\\
 $F_{PL}$     & 12.0               & 10.9                      & 11.06                  & 9.58                     & ~ & ~ &~ & ~ &\\
 FR$^b$       & 0.0125             & 0.0174                    & 0.0043                 & 0.0062                   & ~ & ~ &~ & ~ &\\
 \hline
\end{tabular}
\leftline{$^\dagger$ r024, r13 are combined no-dip (0, 2, 4) and dip (1, 3) phases with IXPE (2-6~keV and 2-4~keV respectively) and NuSTAR data.}
\leftline{$^a$ Parameters values were fixed. $^b$ FR is the ratio between thermal (DBB or kerrbb) and nonthermal (PL or nthComp) model fluxes.}
\leftline{The unit of $N_H$ is $10^{22}~{\rm cm}^{-2}$; $kT_{\rm in}$, $kT_e$, $kT_{bb}$ in keV; $i$ in deg; $D$ in kpc; $M_{BH}$ in $M_\odot$; $\dot{M}_d$ in $10^{18}~g~s^{-1}$; $\dot{m}_d$, $\dot{m}_h$ in $\dot{M}_{Edd}$; $X_s$ in $r_s$.}
\leftline{Fluxes of model components are in $10^{-10}~erg~cm^{-2}~s^{-1}$.}
\leftline{Note: for parameters of kerrbb $1\sigma$ errors and for other model parameters errors with 90\% C.L.}
\label{table_spec}
\end{table*}

\subsection{Broadband Spectral Analysis}
To study the broadband nature of the source in the $2$--$70$~keV energy range, we use combined \textit{IXPE} and \textit{NuSTAR} data. The two 
\textit{NuSTAR} observations that overlap with the \textit{IXPE} observation are used (see Fig.~\ref{fig_lc}). The combined spectra are fitted with 
four sets of phenomenological and physical models: $M1:$ $constant\otimes tbabs(diskbb + powerlaw)$; $M2:$ $constant\otimes tbabs(diskbb + nthComp)$;
$M3:$ $constant \otimes tbabs(kerrbb + powerlaw)$, and $M4:$ $constant \otimes tbabs \otimes TCAF$.
The inclusion of \texttt{diskbb} or \texttt{kerrbb} with \texttt{powerlaw} or 
\texttt{nthComp} in models $M1$, $M2$, and $M3$ is statistically justified using the F-test in XSPEC.
In the top panel of Fig.~\ref{fig_spec}, we show the M1 model fitted spectra with its components. In the four bottom panels, we present the variation 
of $\Delta \chi$ obtained from spectral fits with the aforementioned models. The best-fit model parameters and flux contributions in the $2$-$70$~keV 
band by individual model components are listed in Table~\ref{table_spec}.

During spectral fitting, the hydrogen column density $N_H$ of the interstellar absorption model \texttt{TBabs} was found within 
the range of $(1.55$--$1.96)\times 10^{22}$~cm$^{-2}$ for dataset 1 (IXPE+Nu1), and $(2.51$---$2.71)\times 10^{22}$~cm$^{-2}$ for dataset 2 (IXPE+Nu2). 
To avoid inconsistencies, we fixed $N_H$ at $1.87\times 10^{22}$~cm$^{-2}$ for dataset 1, and at $2.61\times 10^{22}$~cm$^{-2}$ for dataset 2, which 
are M1 model fitted values. 
Furthermore, while fitting with model $M3$, the source distance ($D$) and inclination angle ($i$) in the \texttt{kerrbb} model 
were fixed at $14$~kpc [average value from \citet{Rodriguez11}] and $56^\circ$, respectively. Due to poor error constraints, $i$ was fixed at its 
best-fit value obtained during fitting, which is consistent with the estimation reported by \citet{Rao12}.

The source spectrum is largely dominated by non-thermal components, as evident from the variation of the \texttt{diskbb} and 
\texttt{powerlaw} model components (Fig.~\ref{fig_spec}) and the individual flux contributions from models $M1$--$M3$ 
(Table~\ref{table_spec}). The disk temperature $kT_{\rm in}$ is found in the range of $0.41$--$0.44$~keV, and the photon index $\Gamma$ in the range 
$1.53$--$1.58$. Similar values of $kT_{\rm in}$ and $\Gamma$ are found across the other model fits.

The physical \texttt{kerrbb} model incorporates four source parameters (spin, mass, inclination, and distance) and one accretion 
flow parameter (disk rate). The best-fit \texttt{kerrbb} model yields a BH mass $M_{\rm BH}$ in the range 
$14.6$--$14.9~M_\odot$ and spin parameter $a^* \sim 0.54$.

Similarly, the physical \texttt{TCAF} model \citep{CT95,D15} provides accretion flow parameters: Keplerian disk rate ($\dot{m}_d$ 
in Eddington units $\dot{M}_{\rm Edd}$), sub-Keplerian halo rate ($\dot{m}_h$ in $\dot{M}_{\rm Edd}$), shock location ($X_s$ in Schwarzschild radius $r_s$), 
and shock compression ratio ($R$). It also provides an estimate of $M_{\rm BH}$. A higher halo rate ($\sim 0.8$) compared to the disk rate ($<0.01$), 
with a far shock location and strong shock strength (i.e., high $R$), is observed in both datasets. The TCAF model-fitted $M_{\rm BH}$ lies in the range 
$14.5$r--$15.0~M_\odot$.

Assuming shock oscillation as the origin of the QPOs, \citet{D14} estimated the primary frequency of QPOs ($\nu_{\rm QPO}$) for three BHCs. Using the 
same prescription, we estimated the values of $\nu_{\rm QPO}$ for IGR~J17091-3624 as $0.232\pm0.038$~Hz (for Nu1) and $0.214\pm0.069$~Hz (for Nu2). 
These theoretical estimates are consistent with the observed QPOs in the two \textit{NuSTAR} observations, confirming shock oscillation as the origin 
of the observed $\sim0.21$~Hz QPOs in IGR~J17091-3624.

\begin{figure}
  \centering
    \includegraphics[angle=0,width=8.6cm,keepaspectratio=true]{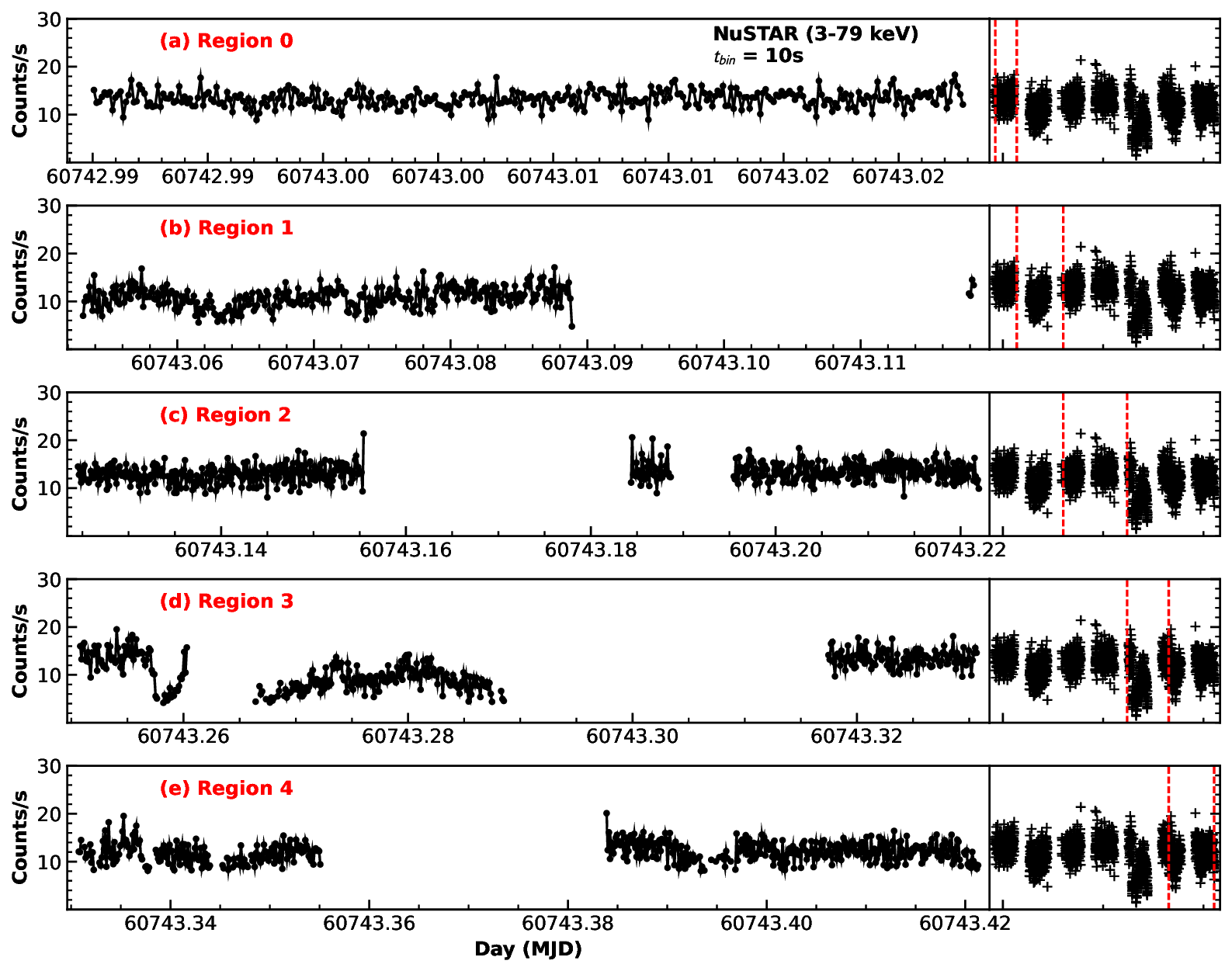}
	\caption{The left panels show light curves of 10 s time bin for five different regions (marked between vertical lines in the right panels) of Nu2.}
\label{fig_lc_rw}
\end{figure}

\begin{figure}
  \centering
    \includegraphics[angle=0,width=8.6cm,keepaspectratio=true]{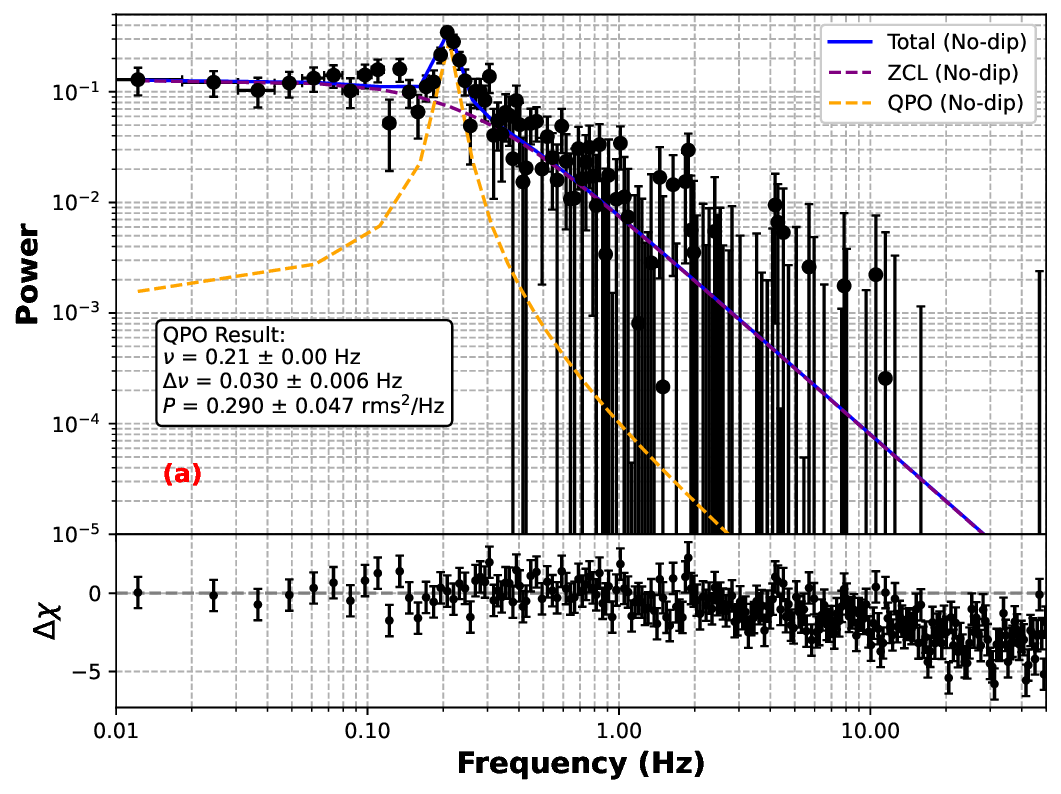}\vskip 0.2cm
    \includegraphics[angle=0,width=8.6cm,keepaspectratio=true]{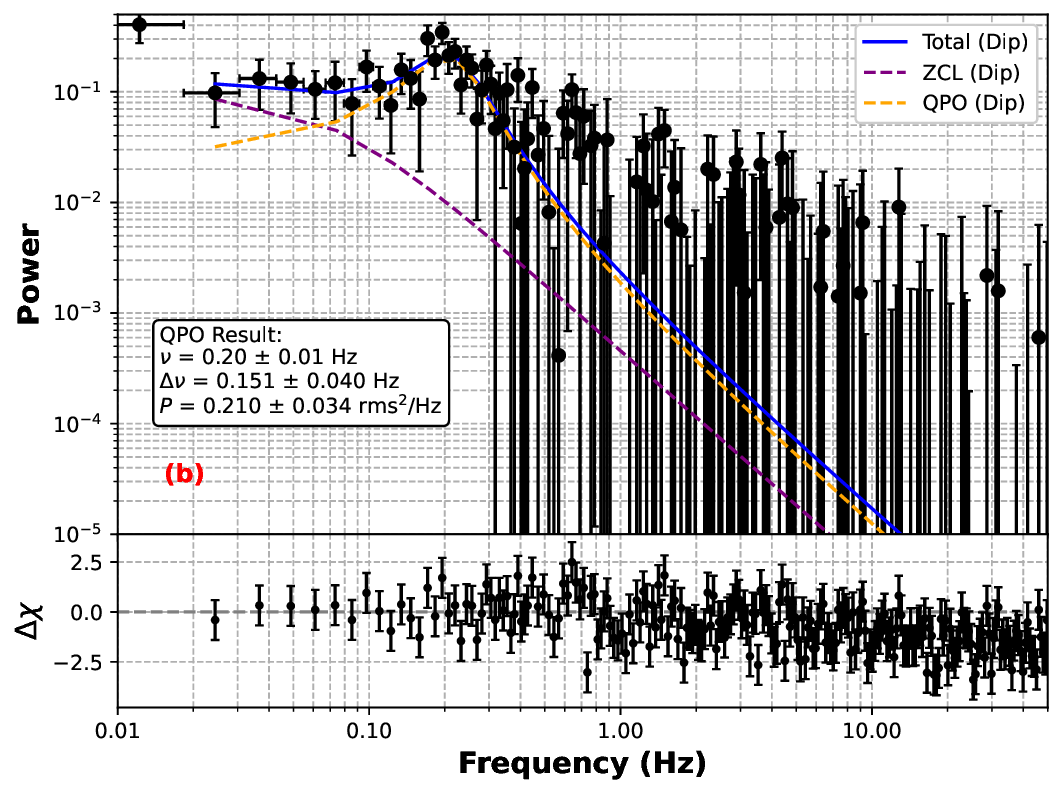}
	\caption{Lorentzian model fitted power density spectrum (PDS) of (a) no-dip (0, 2, 4), and (b) dip (1, 3) regions of Nu2. 
	PDS are fitted with combination of zero-centric (ZCL) qnd QPO-centric Lorentzian models. Inset $\nu$, $\Delta\nu$, and \text{P} 
	mark model fitted QPO frequency, FWHM, and power. Note for fitting dip region PDS, frequency $\nu<0.02$~Hz is not considered.}
\label{fig_pds_rw}
\end{figure}

\subsection{Study in Dip and No-Dip Phases of Nu2}
To understand the nature of the source and the requirement of a higher hydrogen column density ($N_{\rm H}$) in Nu2, we carried out 
detailed temporal and spectral studies of both the dip and no-dip regions as observed in the IXPE and NuSTAR light curves (see Fig.~\ref{fig_lc}). 
Based on the variations in the light curves, the Nu2 observational period is divided into five regions: 0 to 4. Regions 0 
(MJD 60742.9805--60743.0380), 2 (MJD 60743.1215--60743.2366), and 4 (MJD 60743.3287--60743.4213) correspond to no-dip phases (similar to Nu1), 
while regions 1 (MJD 60743.0380--60743.1215) and 3 (MJD 60743.2366--60743.3287) correspond to dip phases. The durations of regions 0 to 4 
are approximately 5, 7, 10, 8, and 8~ks, respectively. We also performed time-dependent polarization analysis in different regions of the Nu2 
observation period. However, the detection values were below the Minimum Detectable Polarization (MDP) threshold (99\%).

\subsubsection{Region-wise Light Curve}
To explore the detailed variation of X-ray intensities across different regions of Nu2, 10~s time-binned light curves over intervals are 
shown in Fig.~\ref{fig_lc_rw}. Although the variability in region 3 (dip phase) roughly resembles the $\lambda$ or $\beta$ classes of IGR~J17091--3624 
or GRS~1915+105 \citep{Altamirano11}, it features a longer low-intensity interval ($\sim200$~s). Other regions appear similar to the $\chi$ class 
\citep{Belloni00}. A more detailed study is required to confirm the classes in the light curves, which will be carried out and published elsewhere.

\subsubsection{Region-wise Power Density Spectrum}
In Fig.~\ref{fig_pds_rw}, we present Lorentzian model-fitted power density spectra (PDS) for the combined dip (regions 1 and 3) and no-dip 
(regions 0, 2, and 4) phases. The QPO sharpness is significantly reduced in the dip phase, as the Q-value ($\nu/\Delta\nu$) drops from 7 to 1.3. 
This indicates a transition from a type-C QPO in the no-dip phase to a type-A QPO in the dip phase.

\subsubsection{Region-wise Spectrum}
The joint IXPE and NuSTAR spectra for the dip and no-dip phases in the combined energy band of 2--70~keV were fitted using four sets of models, 
as described earlier (Table~\ref{table_spec}). Due to the low signal-to-noise ratio in the no-dip and dip phases, we combined 2--6~keV and 2--4~keV IXPE 
data respectively with 3--70~keV NuSTAR data. A significantly higher $N_{\rm H}$ is required in the dip regions ($6.03 \times 10^{22}~\mathrm{cm}^{-2}$) 
compared to the no-dip regions ($2.03 \times 10^{22}~\mathrm{cm}^{-2}$). For reference, during the entire Nu2 observation, the average $N_{\rm H}$ 
was found to be $2.61 \times 10^{22}~\mathrm{cm}^{-2}$, which likely reflects an average of the dip and no-dip values.
Although the photon index ($\Gamma$) did not change significantly, a decrease in nonthermal (PL or \texttt{nthComp}) model fluxes and 
an increase in thermal (DBB or \texttt{kerrbb}) model fluxes during the dip phase indicate spectral softening within the hard state. 
The {\fontfamily{qcr}\selectfont TCAF} model also shows an increase in the Keplerian disk rate ($\dot{m}_d$) during the dip phase.

\section{Discussion and Concluding Remarks}
The class-transitioning Galactic transient BHC IGR~J17091-3624 has recently been observed simultaneously by \textit{IXPE} and 
\textit{NuSTAR} satellites. This is a very interesting source, similar to GRS~1915+105, exhibiting various classes of light curves. 
Here, we present a spectro-polarimetric study of the source using a total of $163$~ks \textit{IXPE} observation and two 
\textit{NuSTAR} observations each of $38$~ks, carried out between March 7--10, 2025 (MJD 60741.31--60744.77).

The IXPE light curve shows three prominent dips on March 9, 2025 (MJD 60743--60743.5), which are also clearly visible in the \textit{NuSTAR} 
light curve (see Fig.~\ref{fig_lc}). A detailed spectral and timing analysis in the dip and no-dip phases of Nu2 observation is done. 

The polarization measurements of IGR~J17091-3624 have been carried out using both model-independent (\texttt{PCUBE} algorithm) 
and model-dependent (spectral fits) methods. The \texttt{PCUBE} algorithm-based analysis revealed a significant polarization 
degree of PD = ($11.3 \pm 2.35$)\% at a PA of $(82.7 \pm 5.95)^\circ$ in the $2$--$8$~keV range, with a confidence level 
exceeding $4\sigma$.

An energy-dependent evolution of the polarization degree is observed, increasing from ($7.27 \pm 2.16$)\% to ($29.9 \pm 8.46$)\% across the three 
IXPE energy bands: $2$--$4$, $4$--$6$, and $6$--$8$~keV (see Fig.~\ref{fig_pdpa}(a--b) and Table~\ref{table_pol}). The highest PD = ($29.9 \pm 8.46$)\% 
is observed at PA = $(88.0 \pm 8.15)^\circ$ in the highest energy band ($6$--$8$~keV). This suggests that, similar to Cyg~X-1, a larger polarization degree 
might be observed in even harder X-ray and $\gamma$-ray bands. Physically, a high PD at higher energies reflects the geometry, emission mechanisms, 
and general relativistic effects experienced by the photons. Since higher-energy photons undergo more inverse-Compton scatterings and originate closer 
to the BH, they experience stronger relativistic effects—especially in the case of rapidly spinning BHs. Therefore, we expect 
IGR~J17091-3624 to host a rapidly spinning BH. The broadband spectral analysis in this work is also used to estimate the spin of the source.

For the model-dependent polarization measurements, we employ two polarimetric models: \texttt{polconst} and \texttt{polpow}, to fit the Stokes 
spectra (PHA1, PHA1Q, and PHA1U) of all three IXPE detector units (DUs) in the $2$--$8$~keV band using the \texttt{XSPEC} software package. 
The \texttt{polconst} model yields a polarization degree (PD) of $(7.92 \pm 1.03)\%$ and a polarization angle (PA) of $(82.16 \pm 3.73)^\circ$.
Similarly, the \texttt{polpow} model provides a PD of $(10.51 \pm 3.17)\%$ at a PA of $(80.56 \pm 8.40)^\circ$. In Fig.~\ref{fig_pdpa}(c),
we present the PD and PA values along with their $1\sigma$ confidence contours as obtained from both models. 
The closer agreement of the \texttt{polpow} model PD value with the model-independent result suggests that the polarization is not constant,
but instead evolves with energy. The \texttt{polpow} model also allows us to estimate the PD and PA values in different energy bands within 
the \textit{IXPE} spectral range of $2$--$8$~keV. The estimated values in the three bands—$2$--$4$, $4$--$6$, and $6$--$8$~keV show a similar 
monotonic increase in PD, consistent with the model-independent analysis (see Table~\ref{table_pol}).

To study the broadband nature and accretion flow properties of the source, the combined \textit{IXPE} and \textit{NuSTAR} spectra are fitted with four 
different sets of models. Both phenomenological and physical models fits indicate a strong dominance of nonthermal flux originating from the `hot' Compton 
cloud or corona (see individual model component fluxes in Tables~\ref{table_pol} \& \ref{table_spec} and Fig.~\ref{fig_spec}). Hence, we infer that during 
the observation period, the source was in the hard state (HS). The presence of a cooler disk ($kT_{\rm in} \sim 0.4$~keV) and a low photon index 
($\Gamma \sim 1.6$) as obtained from models M1 and M2 further supports this spectral classification.

The physical \texttt{kerrbb} model allows the inference of intrinsic source parameters such as BH mass, spin, inclination angle, 
and distance, in addition to the disk accretion rate. By keeping the distance and inclination angle fixed at their reported or best-fit values, we estimate 
the mass and spin of the source. For the two \textit{NuSTAR} datasets, the mass is found to be $14.9^{+0.4}_{-1.3}~M_\odot$ and $14.6^{+4.8}_{-3.4}~M_\odot$, 
and the spin parameter ($a^*$) is estimated as $0.54^{+0.09}_{-0.07}$ and $0.54^{+0.14}_{-0.25}$, respectively. Due to large uncertainties in the second 
dataset, the estimated spin is constrained with weak confidence. Nonetheless, this value is consistent with the earlier estimate of $a^* > 0.53$ by 
Rao \& Vadawale (2012).

The physical \texttt{TCAF} model directly estimates the accretion flow parameters and the BH mass. In both datasets, a strong dominance of the 
sub-Keplerian halo rate over the Keplerian disk rate is observed. A larger shock location and higher compression ratio imply the presence of 
an extended corona with enhanced post-shock densities. The detection of a higher PD is consistent with this coronal geometry. The \texttt{TCAF} 
model fits yield BH masses of $14.4 \pm 1.7~M_\odot$ and $15.0 \pm 3.6~M_\odot$ for datasets 1 and 2, respectively. 
Combining the \texttt{TCAF} and \texttt{kerrbb} model estimates, the most probable BH mass is inferred to be $14.8^{+4.7}_{-3.4}~M_\odot$. 
Additionally, the shock parameters derived from the \texttt{TCAF} model allow for the estimation of the primary QPO frequencies, which are 
consistent with the observed values. This supports the interpretation that the observed QPOs originate from shock oscillations in the accretion flow.

To investigate the physical reason behind the requirement of a higher $N_{\rm H}$ value in Nu2 compared to Nu1, we performed detailed spectral 
and temporal analyses of the dip and no-dip phases of Nu2. The four sets of spectral models clearly indicate an excess of thermal flux relative to 
nonthermal flux during the dip phase, as the flux ratio (FR; defined as the ratio between thermal and nonthermal fluxes) 50\% more in this 
phase (see Table~\ref{table_spec}). The {\fontfamily{qcr}\selectfont TCAF} model also shows an increase in the Keplerian disk accretion rate during 
the dip phase. A significantly higher $N_{\rm H}$ is required to fit the spectrum during the dip phase compared to the no-dip phase of Nu2. This 
possibly caused due to the obscuration of the Compton cloud by wind material accreted from the companion star \citep[see,][]{King12, Athulya23}.
The sharpness (Q-value) of the QPO in the power density spectrum (PDS) is also reduced in the dip state, and the QPO changes from type-C 
in the no-dip phase to type-A in the dip phase. This transition may be attributed to a decrease in the nonthermal hard flux contribution from the 
Compton cloud, whose resonance oscillation is believed to be responsible for the origin of type-C QPOs \citep{C15}.

A brief summary of our findings in this {\it paper} is as follows:

\begin{enumerate}[i)]

\item The spectro-polarimetric study of IGR~J17091-3624 reveals, the detection and detailed significance of the polarization properties of the source.

\item A significant polarization degree at a large polarization angle is estimated from both our model-independent and model-dependent analyses 
	using \textit{IXPE} data.

\item The observed increase in polarization degree with energy suggests the presence of a larger corona, with more energetic photons being emitted 
	from its inner region and experiencing stronger relativistic effects.

\item The broadband spectral study indicates that IGR~J17091-3624 was in the hard state during the observation period, characterized by a dominant 
	contribution of nonthermal photons from the `hot' corona or Compton cloud.

\item The physical \texttt{kerrbb} model estimates the spin of the black hole to be $a^* \sim 0.54$. This high spin value is 
	consistent with the detection of a large polarization degree.

\item The physical \texttt{TCAF} model–fitted shock parameters directly infer the presence of a large corona or Compton cloud. 
	The high dominance of the sub-Keplerian halo accretion rate confirms the prevalence of Comptonizing processes.

\item Both the \texttt{TCAF} and \texttt{kerrbb} models estimate the black hole mass. The combined results 
	yield a mass of $M_{\rm BH} = 14.8^{+4.7}_{-3.4}~M_\odot$.

\item The agreement between the observed QPO frequency and that predicted from the \texttt{TCAF} model–fitted shock parameters 
	supports the interpretation that the observed LFQPO originates from shock oscillations.

\item During the second NuSTAR (Nu2) observational period, both IXPE and NuSTAR light curves exhibit dips. In the spectral analysis, 
      a higher $N_{\rm H}$ is required to fit the broadband ($2$–$79$~keV) IXPE plus Nu2 data. Detailed spectral and timing analysis of the 
      dip and non-dip regions of Nu2 suggest an increase in thermal flux during the dip phase, likely due to obscuration of the Compton cloud 
      by the wind flow accreted from the companion.

\end{enumerate}

\section*{Acknowledgements}

We are thankful to the anonymous referee for his/her valuable suggestions, which helped improve the quality of the paper. 
We also thank Dr. Anuj Nandi of U.R. Rao Satellite Centre, Bengaluru, India, for sharing his expertise.
This work made use of NICER/XTI and NuSTAR/FPMA data supplied by the High Energy Astrophysics Science Archive Research Center (HEASARC) archive.
D.D. acknowledge the visiting research grant of National Tsing Hua University, Taiwan (NSTC NSTC 113-2811-M-007-010). 
H.-K. C. is supported by NSTC of Taiwan under grant NSTC 113-2112-M-007-020.


\end{document}